\documentclass[twocolumn,letterpaper,aps,prc,longbibliography,superscriptaddress,showpacs,nofootinbib,floatfix]{revtex4-1}

\usepackage{graphicx}	
\usepackage{dcolumn}
\usepackage{bm}
\usepackage{xspace}	
\usepackage{amsmath}

\newcommand{\pt}{\mbox{$p_T$}\xspace}

\newcommand{\sqsn}{\mbox{$\sqrt{s_{_{NN}}}$}\xspace}

\newcommand{\dau}{\mbox{$d$$+$Au}\xspace}

\newcommand{\pau}{\mbox{$p$$+$Au}\xspace}

\newcommand{\hau}{\mbox{$^3{\rm He}$$+$Au}\xspace}
\newcommand{\pp}{\mbox{$p$+$p$}\xspace}
\newcommand{\ppb}{\mbox{$p$$+$Pb}\xspace}

\newcommand{\bbceta}{\mbox{$3.0<|\eta|<3.9$}\xspace}

\begin{document}

\title{Measurement of long-range angular correlations and azimuthal 
anisotropies in high-multiplicity $p$$+$Au collisions 
at~$\sqrt{s_{_{NN}}}=200$~GeV}

\newcommand{\abilene}{Abilene Christian University, Abilene, Texas 79699, USA}
\newcommand{\augie}{Department of Physics, Augustana University, Sioux Falls, South Dakota 57197, USA}
\newcommand{\banaras}{Department of Physics, Banaras Hindu University, Varanasi 221005, India}
\newcommand{\barc}{Bhabha Atomic Research Centre, Bombay 400 085, India}
\newcommand{\baruch}{Baruch College, City University of New York, New York, New York, 10010 USA}
\newcommand{\bnlcoll}{Collider-Accelerator Department, Brookhaven National Laboratory, Upton, New York 11973-5000, USA}
\newcommand{\bnlphys}{Physics Department, Brookhaven National Laboratory, Upton, New York 11973-5000, USA}
\newcommand{\caucr}{University of California-Riverside, Riverside, California 92521, USA}
\newcommand{\charlesczech}{Charles University, Ovocn\'{y} trh 5, Praha 1, 116 36, Prague, Czech Republic}
\newcommand{\chonbuk}{Chonbuk National University, Jeonju, 561-756, Korea}
\newcommand{\cns}{Center for Nuclear Study, Graduate School of Science, University of Tokyo, 7-3-1 Hongo, Bunkyo, Tokyo 113-0033, Japan}
\newcommand{\colorado}{University of Colorado, Boulder, Colorado 80309, USA}
\newcommand{\columbia}{Columbia University, New York, New York 10027 and Nevis Laboratories, Irvington, New York 10533, USA}
\newcommand{\czechtech}{Czech Technical University, Zikova 4, 166 36 Prague 6, Czech Republic}
\newcommand{\debrecen}{Debrecen University, H-4010 Debrecen, Egyetem t{\'e}r 1, Hungary}
\newcommand{\elte}{ELTE, E{\"o}tv{\"o}s Lor{\'a}nd University, H-1117 Budapest, P{\'a}zm{\'a}ny P.~s.~1/A, Hungary}
\newcommand{\eszterhazy}{Eszterh\'azy K\'aroly University, K\'aroly R\'obert Campus, H-3200 Gy\"ngy\"os, M\'atrai \'ut 36, Hungary}
\newcommand{\ewha}{Ewha Womans University, Seoul 120-750, Korea}
\newcommand{\fsu}{Florida State University, Tallahassee, Florida 32306, USA}
\newcommand{\gsu}{Georgia State University, Atlanta, Georgia 30303, USA}
\newcommand{\hiroshima}{Hiroshima University, Kagamiyama, Higashi-Hiroshima 739-8526, Japan}
\newcommand{\howard}{Department of Physics and Astronomy, Howard University, Washington, DC 20059, USA}
\newcommand{\ihepprot}{IHEP Protvino, State Research Center of Russian Federation, Institute for High Energy Physics, Protvino, 142281, Russia}
\newcommand{\illuiuc}{University of Illinois at Urbana-Champaign, Urbana, Illinois 61801, USA}
\newcommand{\inrras}{Institute for Nuclear Research of the Russian Academy of Sciences, prospekt 60-letiya Oktyabrya 7a, Moscow 117312, Russia}
\newcommand{\instpasczech}{Institute of Physics, Academy of Sciences of the Czech Republic, Na Slovance 2, 182 21 Prague 8, Czech Republic}
\newcommand{\isu}{Iowa State University, Ames, Iowa 50011, USA}
\newcommand{\jaea}{Advanced Science Research Center, Japan Atomic Energy Agency, 2-4 Shirakata Shirane, Tokai-mura, Naka-gun, Ibaraki-ken 319-1195, Japan}
\newcommand{\jyvaskyla}{Helsinki Institute of Physics and University of Jyv{\"a}skyl{\"a}, P.O.Box 35, FI-40014 Jyv{\"a}skyl{\"a}, Finland}
\newcommand{\kek}{KEK, High Energy Accelerator Research Organization, Tsukuba, Ibaraki 305-0801, Japan}
\newcommand{\korea}{Korea University, Seoul, 136-701, Korea}
\newcommand{\kurchatov}{National Research Center ``Kurchatov Institute", Moscow, 123098 Russia}
\newcommand{\kyoto}{Kyoto University, Kyoto 606-8502, Japan}
\newcommand{\lawllnl}{Lawrence Livermore National Laboratory, Livermore, California 94550, USA}
\newcommand{\losalamos}{Los Alamos National Laboratory, Los Alamos, New Mexico 87545, USA}
\newcommand{\lund}{Department of Physics, Lund University, Box 118, SE-221 00 Lund, Sweden}
\newcommand{\maryland}{University of Maryland, College Park, Maryland 20742, USA}
\newcommand{\mass}{Department of Physics, University of Massachusetts, Amherst, Massachusetts 01003-9337, USA}
\newcommand{\michigan}{Department of Physics, University of Michigan, Ann Arbor, Michigan 48109-1040, USA}
\newcommand{\muhlenberg}{Muhlenberg College, Allentown, Pennsylvania 18104-5586, USA}
\newcommand{\nara}{Nara Women's University, Kita-uoya Nishi-machi Nara 630-8506, Japan}
\newcommand{\natmephi}{National Research Nuclear University, MEPhI, Moscow Engineering Physics Institute, Moscow, 115409, Russia}
\newcommand{\newmex}{University of New Mexico, Albuquerque, New Mexico 87131, USA}
\newcommand{\nmsu}{New Mexico State University, Las Cruces, New Mexico 88003, USA}
\newcommand{\ohio}{Department of Physics and Astronomy, Ohio University, Athens, Ohio 45701, USA}
\newcommand{\ornl}{Oak Ridge National Laboratory, Oak Ridge, Tennessee 37831, USA}
\newcommand{\orsay}{IPN-Orsay, Univ.~Paris-Sud, CNRS/IN2P3, Universit\'e Paris-Saclay, BP1, F-91406, Orsay, France}
\newcommand{\peking}{Peking University, Beijing 100871, People's Republic of China}
\newcommand{\pnpi}{PNPI, Petersburg Nuclear Physics Institute, Gatchina, Leningrad region, 188300, Russia}
\newcommand{\riken}{RIKEN Nishina Center for Accelerator-Based Science, Wako, Saitama 351-0198, Japan}
\newcommand{\rikjrbrc}{RIKEN BNL Research Center, Brookhaven National Laboratory, Upton, New York 11973-5000, USA}
\newcommand{\rikkyo}{Physics Department, Rikkyo University, 3-34-1 Nishi-Ikebukuro, Toshima, Tokyo 171-8501, Japan}
\newcommand{\saispbstu}{Saint Petersburg State Polytechnic University, St.~Petersburg, 195251 Russia}
\newcommand{\seoulnat}{Department of Physics and Astronomy, Seoul National University, Seoul 151-742, Korea}
\newcommand{\stonybrkc}{Chemistry Department, Stony Brook University, SUNY, Stony Brook, New York 11794-3400, USA}
\newcommand{\stonycrkp}{Department of Physics and Astronomy, Stony Brook University, SUNY, Stony Brook, New York 11794-3800, USA}
\newcommand{\tenn}{University of Tennessee, Knoxville, Tennessee 37996, USA}
\newcommand{\titech}{Department of Physics, Tokyo Institute of Technology, Oh-okayama, Meguro, Tokyo 152-8551, Japan}
\newcommand{\tsukuba}{Center for Integrated Research in Fundamental Science and Engineering, University of Tsukuba, Tsukuba, Ibaraki 305, Japan}
\newcommand{\vandy}{Vanderbilt University, Nashville, Tennessee 37235, USA}
\newcommand{\weizmann}{Weizmann Institute, Rehovot 76100, Israel}
\newcommand{\wigner}{Institute for Particle and Nuclear Physics, Wigner Research Centre for Physics, Hungarian Academy of Sciences (Wigner RCP, RMKI) H-1525 Budapest 114, POBox 49, Budapest, Hungary}
\newcommand{\yonsei}{Yonsei University, IPAP, Seoul 120-749, Korea}
\newcommand{\zagreb}{University of Zagreb, Faculty of Science, Department of Physics, Bijeni\v{c}ka 32, HR-10002 Zagreb, Croatia}
\affiliation{\abilene}
\affiliation{\augie}
\affiliation{\banaras}
\affiliation{\barc}
\affiliation{\baruch}
\affiliation{\bnlcoll}
\affiliation{\bnlphys}
\affiliation{\caucr}
\affiliation{\charlesczech}
\affiliation{\chonbuk}
\affiliation{\cns}
\affiliation{\colorado}
\affiliation{\columbia}
\affiliation{\czechtech}
\affiliation{\debrecen}
\affiliation{\elte}
\affiliation{\eszterhazy}
\affiliation{\ewha}
\affiliation{\fsu}
\affiliation{\gsu}
\affiliation{\hiroshima}
\affiliation{\howard}
\affiliation{\ihepprot}
\affiliation{\illuiuc}
\affiliation{\inrras}
\affiliation{\instpasczech}
\affiliation{\isu}
\affiliation{\jaea}
\affiliation{\jyvaskyla}
\affiliation{\kek}
\affiliation{\korea}
\affiliation{\kurchatov}
\affiliation{\kyoto}
\affiliation{\lawllnl}
\affiliation{\losalamos}
\affiliation{\lund}
\affiliation{\maryland}
\affiliation{\mass}
\affiliation{\michigan}
\affiliation{\muhlenberg}
\affiliation{\nara}
\affiliation{\natmephi}
\affiliation{\newmex}
\affiliation{\nmsu}
\affiliation{\ohio}
\affiliation{\ornl}
\affiliation{\orsay}
\affiliation{\peking}
\affiliation{\pnpi}
\affiliation{\riken}
\affiliation{\rikjrbrc}
\affiliation{\rikkyo}
\affiliation{\saispbstu}
\affiliation{\seoulnat}
\affiliation{\stonybrkc}
\affiliation{\stonycrkp}
\affiliation{\tenn}
\affiliation{\titech}
\affiliation{\tsukuba}
\affiliation{\vandy}
\affiliation{\weizmann}
\affiliation{\wigner}
\affiliation{\yonsei}
\affiliation{\zagreb}
\author{C.~Aidala} \affiliation{\michigan} 
\author{Y.~Akiba} \email[PHENIX Spokesperson: ]{akiba@rcf.rhic.bnl.gov} \affiliation{\riken} \affiliation{\rikjrbrc} 
\author{M.~Alfred} \affiliation{\howard} 
\author{V.~Andrieux} \affiliation{\michigan} 
\author{K.~Aoki} \affiliation{\kek} 
\author{N.~Apadula} \affiliation{\isu} 
\author{H.~Asano} \affiliation{\kyoto} \affiliation{\riken} 
\author{C.~Ayuso} \affiliation{\michigan} 
\author{B.~Azmoun} \affiliation{\bnlphys} 
\author{V.~Babintsev} \affiliation{\ihepprot} 
% ...
\collaboration{PHENIX Collaboration} \noaffiliation

\date{\today}


\begin{abstract}

We present the first measurements of long-range angular correlations and 
the transverse momentum dependence of elliptic flow $v_2$ in 
high-multiplicity \pau collisions at \sqsn = 200 GeV. A comparison of 
these results with previous measurements in high-multiplicity \dau and 
\hau collisions demonstrates a relation between $v_2$ and the initial 
collision eccentricity $\varepsilon_2$, suggesting that the observed 
momentum-space azimuthal anisotropies in these small systems have a 
collective origin and reflect the initial geometry. Good agreement is 
observed between the measured $v_2$ and hydrodynamic calculations for 
all systems, and an argument disfavoring theoretical explanations based 
on initial momentum-space domain correlations is presented. The set of 
measurements presented here allows us to leverage the distinct intrinsic 
geometry of each of these systems to distinguish between different 
theoretical descriptions of the long-range correlations observed in 
small collision systems.

\end{abstract}


\maketitle

\section{Introduction}

The azimuthal momentum anisotropy of particle emission relative to the 
participant plane of the collision, as quantified by the Fourier 
coefficients $v_n$ of the final state particle yield, has long been 
considered evidence for the formation of a strongly interacting, 
fluid-like quark-gluon plasma (QGP) in A+A 
collisions~\cite{Heinz:2013th}. Viscous hydrodynamics supports a picture 
in which the initial spatial distribution in energy density, both from 
intrinsic geometry and fluctuations, is propagated into the final state 
as anisotropies in momentum space. The success of hydrodynamics in 
describing various bulk observables of the QGP has lent credence to the 
notion of hydrodynamic flow as the main driver of the $v_{n}$ signal in 
heavy A+A collisions.

However, recent analyses of \dau and \hau collisions at \sqsn = 200 
GeV~\cite{PhysRevLett.111.212301,Adare:2014keg,Adare:2015ctn,Adamczyk:2014fcx} 
at the Relativistic Heavy-Ion Collider (RHIC), and \ppb at \sqsn = 5.02 
TeV, and $p$$+$$p$ collisions at \sqsn = 2.76, 5.02, 7, and 13 
TeV~\cite{alice_long_2013,atlas_observation_2012,cms_observation_2012,Khachatryan:2015lva,Aad:2015gqa,Khachatryan:2010gv,Khachatryan:2016txc} 
at the Large Hadron Collider (LHC) have demonstrated the existence of 
the same kind of azimuthal anisotropy signals commonly interpreted as 
evidence of collective behavior in larger systems. Notably, a feature 
known as \textit{the ridge} has been observed, consisting of a near-side 
(i.e., at small relative azimuth) enhancement in the long-range (i.e., 
at large relative pseudorapidity) azimuthal two-particle correlation. 
From these correlations, substantial elliptic ($v_2$), and triangular 
($v_3$) flow coefficients have been measured in these systems.

Although these observations seem to support the idea of QGP formation in 
small systems, it is not clear that hydrodynamic expansion would 
translate initial geometry into final state momentum anisotropy in this 
regime, where the formed medium is expected to be short-lived. Other 
explanations have been put forth, including initial state effects from 
glasma diagrams~\cite{dusling_azimuthal_2012}, color 
recombination~\cite{Ortiz:2013yxa}, and partonic scattering in transport 
models~\cite{bzdak_elliptic_2014,Ma:2014pva,Koop:2015wea}. 
Transport model calculations, as well as those from hydrodynamics, 
involve the translation of initial geometry into momentum space via 
final state interactions. Transport models describe interactions between 
well defined particles in kinetic theory, while hydrodynamics involves 
fluid elements. In contrast, glasma diagrams take momentum-space domains 
as a starting point, resulting in momentum correlations without any 
final-state interactions. In this initial momentum-space domain picture, 
the correlations averaged over the event should become weaker in going 
from \pau, to \dau, to \hau as the average is taken over a larger number 
of domains, thus diluting the strength of the correlation effect. There 
is no direct correspondence with the initial geometric eccentricity in 
this picture. A key experimental test to resolve the issue consists in 
varying the initial geometry of the system to analyze the extent to 
which it carries into the final state~\cite{nagle_exploiting_2013}.

The PHENIX collaboration has actively pursued this course of study by 
analyzing data from intrinsically elliptic 
(\dau)~\cite{PhysRevLett.111.212301,Adare:2014keg} and triangular 
(\hau)~\cite{Adare:2015ctn} collision systems at \sqsn = 200 GeV. 
Viscous hydrodynamics followed by a hadron cascade has been found to 
accurately reproduce the measured 
$v_n$~\cite{Romatschke:2015gxa,adare_measurement_2014,PhysRevLett.111.212301,Adare:2015ctn,Bozek:2015qpa} 
for these systems.

This article completes the above suite of studies by presenting 
two-particle correlations and the transverse momentum (\pt) dependence 
of $v_2$ for central \pau collisions at \sqsn = 200 GeV. In small system 
collisions, the term \emph{central} refers to events with 
high-multiplicity and the correlation with actual impact parameter is 
weak. These results are compared to those from \dau and \hau collisions, 
as well as to available theoretical calculations. We apply the same 
analysis procedure to all three systems in the same centrality class, to 
provide a controlled comparison from which to draw conclusions.

\section{Methods}

A detailed description of the PHENIX detector can be found in 
Refs.~\cite{Adcox2003469,Aidala:2013vna}. For this analysis, charged 
particles were reconstructed with the two central arm spectrometers, 
consisting of drift chambers and multi-wire proportional pad 
chambers (PC), each covering $|\eta|<0.35$ in pseudorapidity and $\pi/2$ 
in azimuth. Drift chamber tracks are matched to hits in the third 
(outermost) layer of the PC, thus limiting the contribution of tracks 
from decays and photon conversions. The beam-beam counters (BBC) 
comprise two arrays of 64 quartz radiator \v{C}erenkov detectors, 
located longitudinally $\pm$1.44 m away from the center of the 
interaction region (IR), covering \bbceta and 2$\pi$ in azimuth. The 
forward vertex detector (FVTX) is a silicon detector comprising two 
identical end-cap assemblies symmetrically located in the longitudinal 
direction around the IR, covering the pseudorapidity range $1.0 < |\eta| 
< 3.0$. It uses hit clusters to detect charged particles with an 
efficiency greater than 95\%. The arms of the BBC and FVTX in the 
Au-going direction (i.e., $\eta < 0$) are designated as the \emph{south} 
arms and styled BBC-S and FVTX-S, respectively. We use the south arm of 
each of these two detectors to determine the flow event plane. In 
addition, the $z$-vertex of the collision is found using event timing 
information from both arms of the BBC. In this analysis, a $\pm$10 cm 
cut on the collision $z$-vertex was applied. We compare \pau correlation 
functions with those measured in p+p, as described in detail in 
Ref.~\cite{Adare:2015ctn}.

The \pau data set for this analysis was collected during the 2015 
data-taking run at RHIC. It comprises 0.84 billion minimum bias (MB) 
triggered events and 1.4 billion high-multiplicity (HM) triggered 
events. The MB trigger is defined as a coincidence in the same event 
between the BBC detectors~\cite{Allen2003549} in the Au-going and 
$p$-going directions, requiring at least one photomultiplier tube (PMT) 
firing in each; in this way 84$\pm$4\% of the total inelastic \pau cross 
section is captured. The HM trigger is based on the MB trigger, but 
imposes the additional requirement of more than 35 photomultiplier tubes 
firing in the BBC-S. Events that satisfy this trigger condition 
correspond roughly to the 5\% most central event class. The use of this 
trigger allows us to increase our central event sample size by a factor 
of 25.

\begin{figure}[htbp]
  \includegraphics[width=1.1\linewidth]{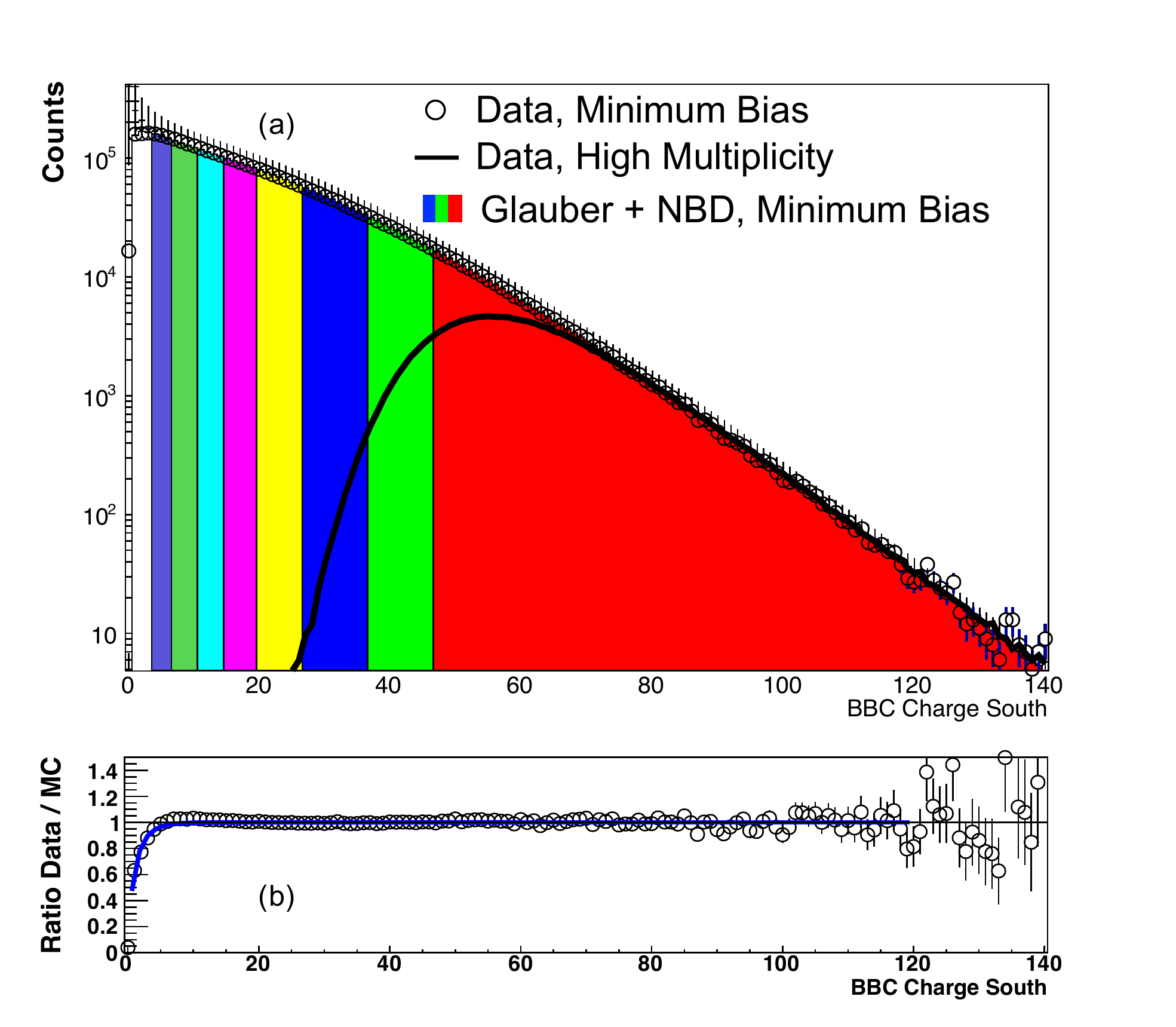}
\caption{(a) BBC-S charge measured in real data from MB (open 
circles) and high multiplicity (solid line) events, where the 
latter distribution has been scaled down by the respective 
trigger prescale factor. The Glauber + NBD calculation is shown 
as [black] crosses.  The shaded histogram [colored areas] 
correspond to the centrality classes for MB events from left to right 
of 0\%--5\%, 5\%--10\%, 10\%--20\%, 20\%--30\%, 30\%--40\%, 
40\%--50\%, 50\%--60\%, 60\%--70\%, and 70\%--88\%.  (b) Ratio 
of real data to the Glauber + NBD calculation for MB events. 
The blue line is a fit to the trigger efficiency turn-on curve.
}
\label{fig:figure0}
\end{figure}

In this analysis, we select the 0\%--5\% most central \pau events, where 
centrality classes are defined by the percentiles of the total 
multiplicity measured in the BBC-S for MB events, following the 
procedure documented in Ref.~\cite{bbc}. Fig.~\ref{fig:figure0}(a) shows 
the measured distribution of BBC-S charge for the MB and HM trigger 
event samples, where the latter has been scaled to match the MB 
distribution. We model the BBC-S charge deposition using a Monte Carlo 
Glauber model with fluctuations following a negative binomial 
distribution. The resulting distribution is shown as a histogram, with 
the colored areas representing various centrality classes. 
Fig.~\ref{fig:figure0}(b) shows the ratio of the measured distribution 
to the MC Glauber calculation for MB events. The inefficiency observed 
below 10 units of charge indicates the MB trigger turn-on.

The initial geometry of events in various centrality selections is 
characterized using a standard Monte Carlo Glauber approach, where 
nucleon coordinates are smeared by a two-dimensional Gaussian of width 
$\sigma = 0.4$ fm. In this model, initial state eccentricity 
$\varepsilon_2$ is computed from initial Gaussian-smeared nucleon 
coordinates, as shown in Eq.~\ref{eqn:epsilon2}.

\begin{equation}
\label{eqn:epsilon2}
\varepsilon_2 = \frac{\sqrt{\langle r^2\cos (2\phi)\rangle ^2 + \langle r^2\sin (2\phi) \rangle ^2}}{\langle r^2 \rangle}
\end{equation}

\begin{table}
\caption{Geometric characterization of small system collisions at 
\sqsn = 200 GeV in the 0\%--5\% centrality class, using Monte Carlo Glauber 
with nucleon coordinates smeared by a two-dimensional Gaussian of width 
$\sigma=0.4$ fm.}
\begin{ruledtabular}  \begin{tabular}{cccc}
\label{table_geometry}
 & \pau & \dau & \hau \\
\hline
 $\langle N_{\rm coll} \rangle$ & $9.7\pm 0.6$ & $18.1\pm 1.2$ & $26.1\pm 2.0$ \\
 $\langle N_{\rm part} \rangle$ & $10.7\pm 0.6$ & $17.8\pm 1.2$ & $25.1\pm  1.6$ \\ 
 Glauber $\langle \varepsilon_2 \rangle$ & $0.23\pm 0.01$ & $0.54\pm 0.04$ & $0.50\pm 0.02$
\end{tabular}  \end{ruledtabular}
\end{table}

In the above equation, $r$ is the radial nucleon position relative to 
the centroid of the participants, and $\phi$ is the nucleon azimuthal 
angle. The results of this Glauber characterization of the initial 
geometry are shown in Table~\ref{table_geometry}. The quantities 
characterizing the event geometry are the same within uncertainties for 
both the MB and HM event samples.

\section{Results}

Long-range angular correlations are constructed between charged tracks 
in the PHENIX central arms at a given \pt, and charge deposited in the 
BBC-S PMTs, for central \pau collisions. The distribution of these 
track-PMT pairs is constructed over relative azimuth as given in 
Eq.~\ref{eq31}, with the normalized correlation function given by 
Eq.~\ref{eq:def_corr_function}, following 
Ref.~\cite{PhysRevLett.115.142301}:

\begin{eqnarray}
  S(\Delta\phi,p_{T})=
  \frac{ d(w_{{\rm PMT}} N^{{\rm track}(p_{T}){\rm - PMT}}_{{\rm Same \; event}}) }{ d\Delta\phi}, & &
\label{eq31} \\
  C(\Delta\phi,p_{T}) =
          \frac{S(\Delta\phi,p_{T})}{M(\Delta\phi,p_{T})} \:
          \frac{\int_{0}^{2\pi} M(\Delta\phi,p_{T}) \, d\Delta\phi}{\int_{0}^{2\pi} S(\Delta\phi,p_{T}) \, d\Delta\phi}. & &
  \label{eq:def_corr_function}
\end{eqnarray}

The weights $w_{{\rm PMT}}$ for each pair correspond to the charge in 
the PMTs comprised in that particular pair. The signal distribution $S$ 
is constructed from pairs in the same event. The mixed distribution $M$ 
is constructed using pairs from different events in the same centrality 
class and collision vertex bin. Ten equally sized bins are used within 
the range of $|z|<$ 10 cm in the event mixing.

\begin{figure*}[htbp]
  \includegraphics[width=0.998\linewidth]{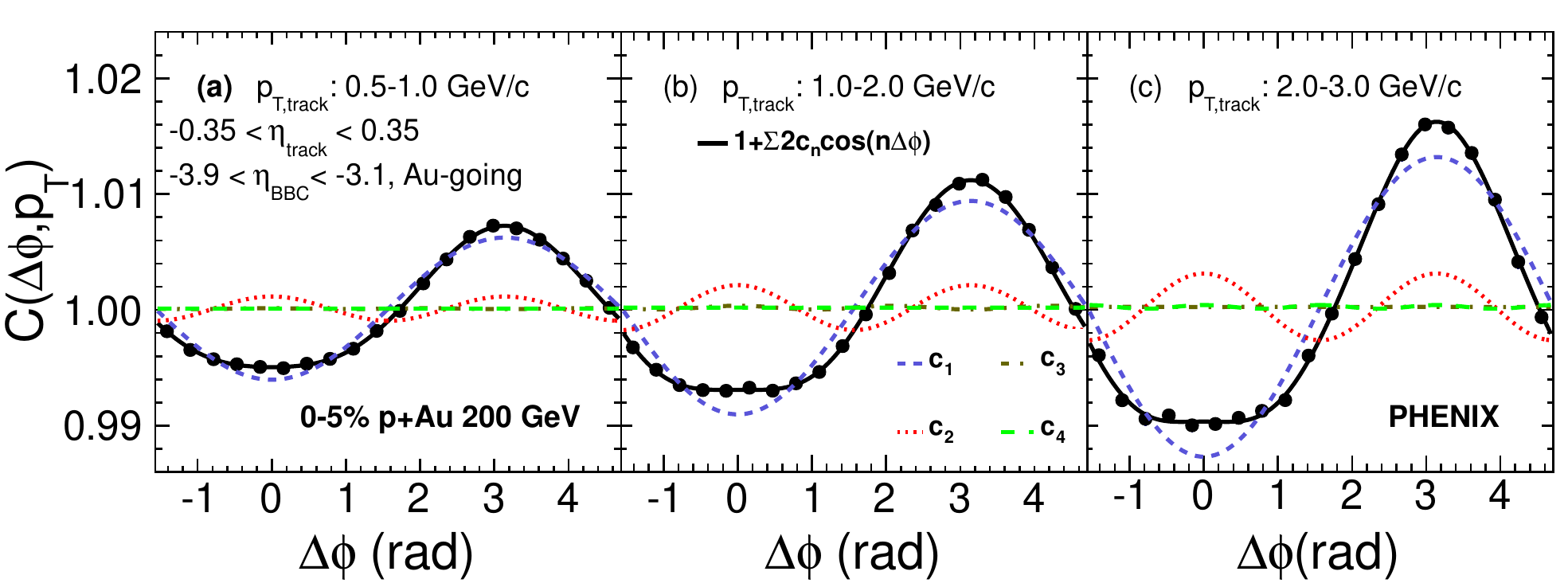}
\caption{Long-range angular correlations $C(\Delta\phi,p_{T})$ 
constructed with central arm tracks and BBC-S PMT pairs, in 0\%--5\% 
central \pau collisions at \sqsn~=~200~GeV. From left to right, 
correlations are shown for various track \pt categories: (a) 0.5--1.0 
GeV/$c$, (b) 1.0--2.0 GeV/$c$, and (c) 2.0--3.0~GeV/$c$. We fit each 
correlation with a four-term cosine Fourier series. The harmonic $c_1$ 
is shown as a short-dashed line; $c_2$, as a dotted line; $c_3$, as a 
dash-dot line; $c_4$, as a long-dashed line. The total fit is shown as a 
solid line.}
\label{fig:figure1}
\end{figure*}

\begin{figure}[htbp]
  \includegraphics[width=1.0\linewidth]{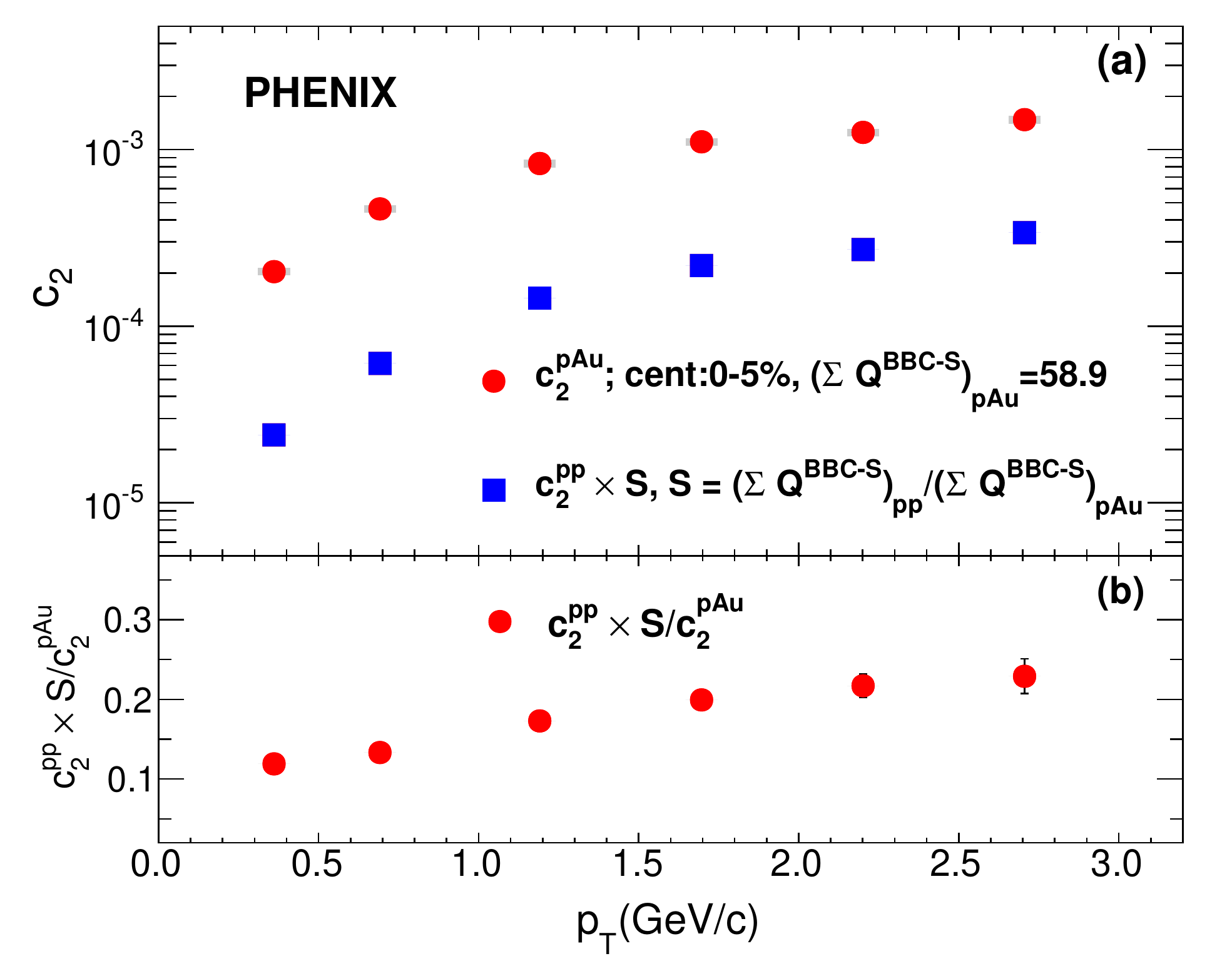}
\caption{(a) The second order harmonic coefficients $c_2(p_T)$ for long 
range angular correlations in 0\%--5\% \pau collisions, as well as for 
MB \pp collisions. The latter are scaled down by the factor 
$\left( \sum Q^{\rm BBC-S} \right)_{p+p} / \left( 
\sum Q^{\rm BBC-S} \right)_{p{\rm Au}}$. (b)~The ratio of the 
two harmonics is plotted with the corresponding statistical errors.
}
\label{fig:figure2}
\end{figure}

The resulting correlation functions for three track \pt selections are 
shown in Fig. ~\ref{fig:figure1}. Each one is fit with a four-term 
cosine Fourier series, $C(\Delta\phi) = 
\sum_{n=1}^{4}2c_n\cos(n\Delta\phi)$. The magnitude of the second 
harmonic $c_{2}$ as a function of \pt is shown with red circles in 
Fig.~\ref{fig:figure2} panel (a). The contribution of elementary 
processes (e.g., jet fragmentation, resonance decays, and momentum 
conservation effects) to the measured $c_2$ in \pau can be estimated 
quantitatively using previously published $c_2$ data from \pp at the 
same collision energy~\cite{Adare:2015ctn}, scaled down by an 
appropriate factor to account for the higher multiplicity in \pau. We 
choose the scale factor to be the ratio of the total charge deposited in 
the BBC-S (i.e., $Q^{{\rm BBC-S}}$) in \pp relative to \pau, as shown 
in Eq.~\ref{eq:dilute}, because we can think of a $p$$+$Au event as the 
superposition of $N$ independent nucleon-nucleon collisions, where the 
correlation strength from a single collision scales inversely with $N$.

\begin{equation}
c_{2}^{p{\rm Au \; elementary}}(p_{T}) \simeq c_{2}^{p+p}(p_{T})
\frac{\left( \sum Q^{\rm BBC-S} \right)_{p+p}}
     {\left( \sum Q^{\rm BBC-S} \right)_{p{\rm Au}}}.
\label{eq:dilute}
\end{equation}

The scaled down reference $c_{2}$ is shown as blue squares in 
Fig.~\ref{fig:figure2}, panel (a). The ratio of $c_2$ in the scaled-down 
\pp reference to \pau is shown in panel (b). From this ratio, it can be 
seen that the relative correlation strength in \pau from elementary 
processes is at most 23\% at the highest \pt.  Because this procedure 
constitutes an approximation to quantify the nonflow correlation 
strength, which may be affected by other factors not considered in this 
analysis, we do not subtract it from the total signal, treating it 
instead as a source of systematic uncertainty. Even though the \pau and 
the \pp baseline data were collected in different years, where potential 
changes in detector performance could affect our results, we verified 
that using \pp data from various run periods has an effect of at most 
3\% on the calculated nonflow contribution.

It is noteworthy that, unlike in \dau~\cite{adare_measurement_2014} and 
\hau~\cite{PhysRevLett.115.142301} collisions at the same centrality, 
the long-range angular correlations in \pau do not exhibit a discernible 
near-side peak, yet possess a nonnegligible second harmonic component. 
The nonflow contribution from elementary processes and momentum 
conservation becomes more dominant as the system size and particle 
multiplicity decrease. This results in a larger $|c_1|$ and thus a 
smaller $|c_2/c_1|$ ratio, and hence in a less discernible near-side 
peak in \pau.

Having quantified the strength of the correlations from elementary 
processes, we determine the second Fourier coefficient $v_2$ of the 
single-particle azimuthal distributions, which is typically associated 
with collective elliptic flow, using the event plane method as described 
in Ref.~\cite{Poskanzer:1998yz}. Namely, we measure

\begin{equation}
v_{2}(p_{T}) = \frac{\langle \cos 2(\phi_{{\rm Particle}}(p_{T})-\Psi^{{\rm FVTX-S}}_{2})\rangle}{{\rm Res}(\Psi^{{\rm FVTX-S}}_{2})}
\end{equation}
for charged hadrons at midrapidity, where the second order event plane 
$\Psi^{{\rm FVTX-S}}_{2}$ is determined for every event using the 
FVTX-S detector. Its resolution ${\rm Res}(\Psi_{2})$ is computed using 
the standard three-subevent method~\cite{Poskanzer:1998yz}, correlating 
measurements in the BBC-S, FVTX-S, and the central arms. This results in 
${\rm Res}(\Psi^{{\rm FVTX-S}}_{2})$ = 0.171. It is also possible to 
measure the event plane using the BBC-S. In that case, we obtain a lower 
resolution ${\rm Res}(\Psi^{{\rm BBC-S}}_{2})$ = 0.062, and $v_2$ 
values that differ from the FVTX-S measurement by approximately 3\%. The 
very good agreement of $v_2$ measured using the BBC-S and FVTX-S 
event planes is interesting, because the pseudorapidity gaps relative to 
the midrapidity tracks are $|\Delta\eta| > 2.65$ and $|\Delta\eta| > 
0.65$, respectively.

The main sources of systematic uncertainty in the $v_2(p_T)$ measurement 
are: (1) track background from photon conversion and weak decays, whose 
magnitude we determine at 2\% relative to the measured $v_2$ by varying 
the spatial matching windows in the PC3 from 3$\sigma$ to 2$\sigma$; (2) 
Multiple collisions per bunch crossing (i.e., event pile-up) that are 
observed to occur at an average rate of 8\% in the 0\%--5\% central $p$$+$Au 
collisions. Low luminosity and high-luminosity subsets of the data were 
analyzed separately and the systematic uncertainty in the $v_2(p_T)$ 
value is determined to be asymmetric $^{+4\%}_{-0\%}$, because the $v_2$ 
values were found to decrease in the events that contain a larger 
fraction of pile-up; (3) Non-flow correlations from elementary processes 
that enhance the $v_2$ values, whose contribution we estimate from 
Fig.~\ref{fig:figure2}, assigning a \pt-dependent asymmetric uncertainty 
with a maximum value of $^{+0}_{-23}\%$ for the highest \pt bin. This 
can be compared to the corresponding $^{+0}_{-9}\%$~\cite{Adare:2014keg} 
and $^{+0}_{-7}\%$~\cite{Adare:2015ctn} systematic uncertainties in \dau 
and \hau collisions, respectively; (4) The asymmetry between the east 
($\pi/2 < \phi < 3\pi/2$) and west ($-\pi/2 < \phi < \pi/2$) acceptance 
of the detectors due to an offset of 3.6 mrad between the colliding 
beams and the longitudinal axis of PHENIX, necessary for running \pau at 
the same momentum per nucleon. We applied a corresponding 
counter-rotation to every central arm track and detector element in the 
FVTX and BBC, which were also reweighted to restore their uniformity in 
azimuth. We assign a value of 5\% for this systematic uncertainty by 
taking the difference of $v_2$ as measured independently in the east and 
the west arms after applying the above corrections; (5) The difference 
in the $v_2(p_T)$ values when measured independently using the BBC-S and 
FVTX event planes, which we observe to differ by $\pm$3\%.

\label{s:sys}
\begin{table}[htbp]
\caption{
Systematic uncertainties given as a percent of the $v_2$ measurement. 
Note that the nonflow contribution is \pt dependent and the value here 
quoted corresponds to the highest measured \pt.}
\begin{ruledtabular}  \begin{tabular}{ccc}
      Source& Systematic Uncertainty & Type \\ \hline
      Track Background &2.0\%& A\\ 
      Event Pile-up    &$^{+4}_{-0}\%$& B\\
      Non-Flow    &$^{+0}_{-23}\%$& B\\
      Beam Angle &5.0\%& C\\  
      Event-Plane Detectors & 3\% & C\\
    \end{tabular} \end{ruledtabular}
\label{t:sys}
 \end{table}

Table~\ref{t:sys} summarizes of all these systematic
uncertainties, categorized by type:

(A) point-to-point uncorrelated between $p_T$ bins,

(B) point-to-point correlated between $p_T$ bins,

(C) overall normalization uncertainty in which all points are scaled by 
the same multiplicative factor.

The resulting $v_2$ measurement for \pau, compared to 
\dau~\cite{Adare:2014keg} and \hau~\cite{Adare:2015ctn} in the same 
0\%--5\% centrality class, is shown in Fig.~\ref{fig:figure3}. The \dau 
data, as presented in Ref.~\cite{Adare:2014keg}, did not include 
nonflow contributions in its systematic uncertainties, which are now 
accounted for in the systematics shown in Fig.~\ref{fig:figure3}. In all 
cases, there is a substantial $v_2$ that rises with \pt. It is notable 
that the $v_2$ values for \dau and \hau are consistent within 
uncertainties, as are their eccentricities $\varepsilon_2$ listed in 
Table I. The \pau collisions have a significantly lower $v_2$ and a 
correspondingly lower calculated $\varepsilon_2$. At the same time, the 
ordering of $v_2$ from \pau, to \dau, to \hau also follows the expected 
increasing order of particle multiplicity. In the case of \dau and \hau, 
for the 0\%--5\% most central events, the published values for 
midrapidity charged particle density are $dN_{ch}/d\eta$ = 20.8 $\pm$ 
1.5 and 26.3 $\pm$ 1.8, respectively~\cite{Adare:2015bua}. This quantity 
has not yet been measured in \pau collisions.

\section{Discussion}

\begin{figure}[htbp]
  \includegraphics[width=1.0\linewidth]{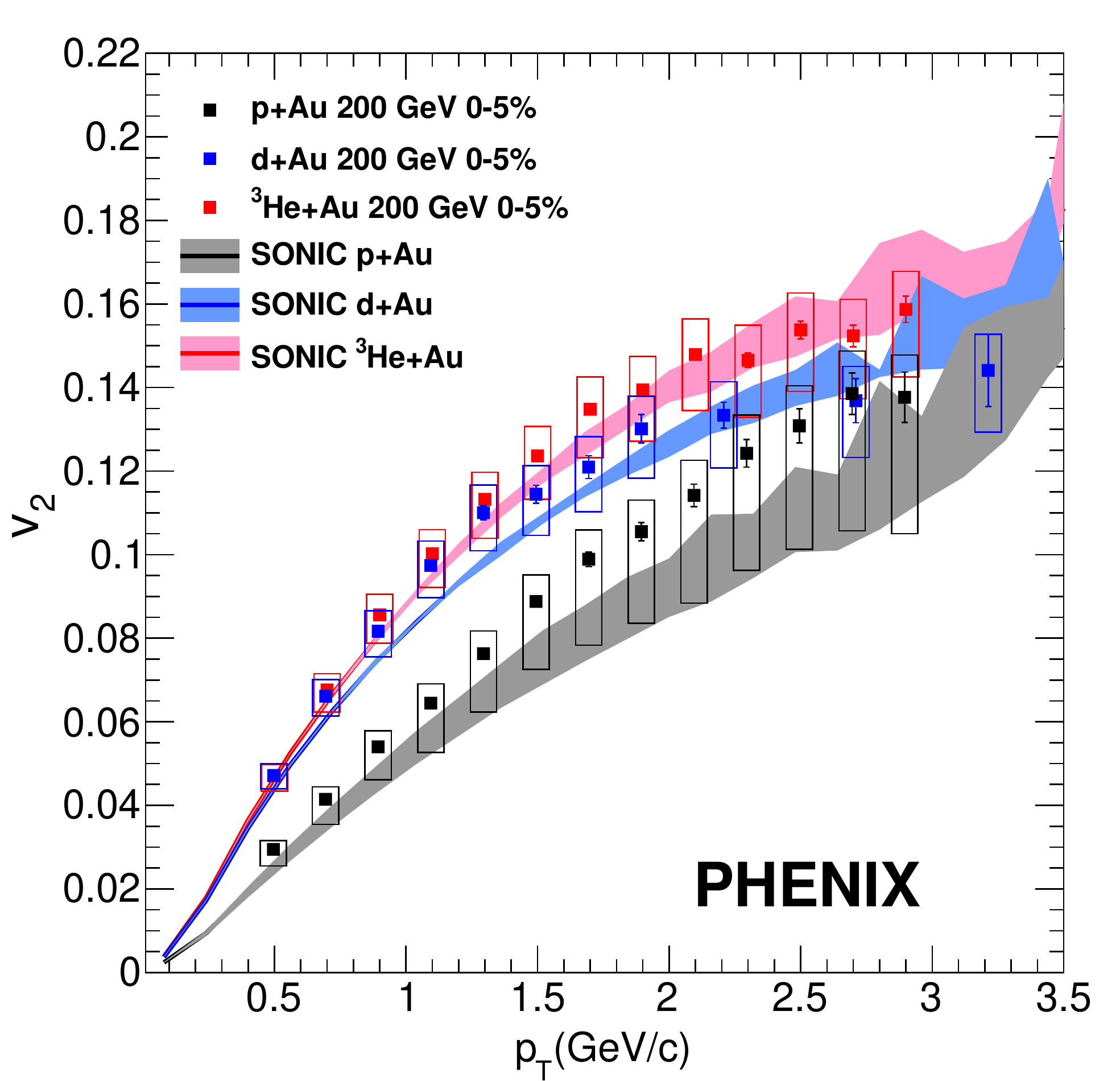}
\caption{$v_2$ of charged hadrons within $|\eta| <$ 0.35 in 0\%--5\% 
(bottom [gray] curve) \pau, (middle [blue] curve) \dau, and 
(top [red] curve) \hau central collisions, compared to hydrodynamic calculations 
using the \textsc{sonic} model, matched to the same 
multiplicity as the data. Note that the data points shown include 
nonflow contributions, whose estimated magnitude is accounted for in 
the asymmetric systematic uncertainties.}
\label{fig:figure3}
\end{figure}

Also shown in Fig. ~\ref{fig:figure3} are $v_2$ calculations for each 
system from the \textsc{sonic} hydrodynamic model~\cite{Habich:2014jna}, 
which incorporates standard Monte Carlo Glauber initial conditions 
followed by viscous hydrodynamics with $\eta/s=0.08$, and a transition 
to a hadronic cascade at $T=$ 170 MeV. It is notable that these 
calculations for each system are matched to the charged particle density 
at midrapidity, with the exact values for 0\%--5\% centrality of 10.0, 
20.0, and 27.0, for \pau, \dau, and \hau collisions, 
respectively~\cite{Habich:2014jna}. Again, note that $dN_{ch}/d\eta$ has 
not been measured for \pau, and that the value of 10.0 was extrapolated 
from measurements in the other two systems~\cite{Habich:2014jna}. We 
thus see that the calculation includes both the geometry-related change 
in eccentricity and the relative collision multiplicity. In all cases, a 
good agreement is seen within uncertainties between the data and the 
calculation. These observations strongly support the notion of initial 
geometry, coupled to the hydrodynamic evolution of the medium as a valid 
framework to understand small system collectivity.

\begin{figure}[htbp]
  \includegraphics[width=1.0\linewidth]{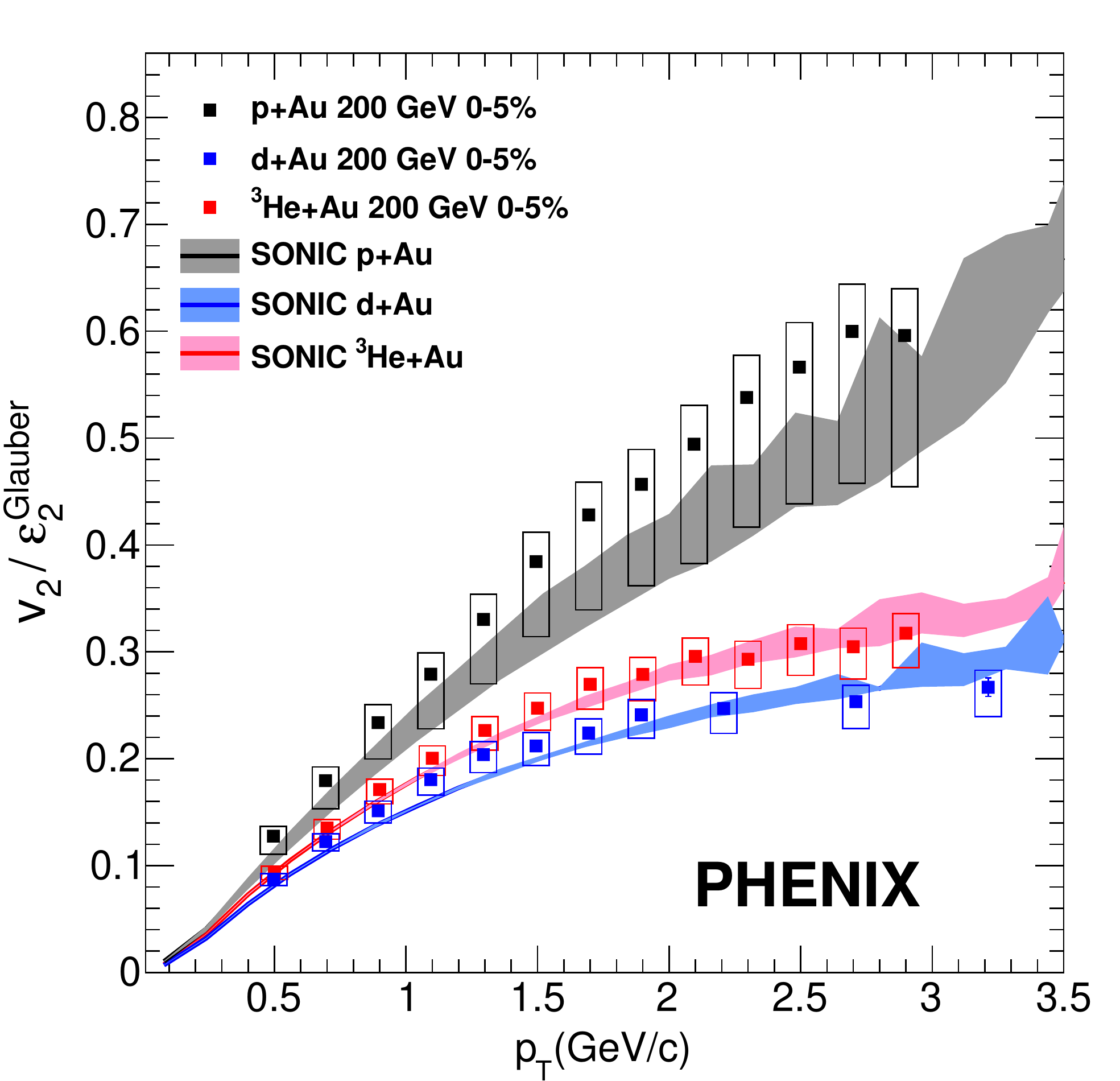}
\caption{$v_2$ of charged hadrons within $|\eta| <$ 0.35 in 0\%--5\% 
(top [gray] curve) \pau, (center [red] curve) \dau and 
(bottom [blue] curve) \hau central collisions, divided by their 
corresponding eccentricity $\varepsilon_2$ from Glauber calculations, 
compared to \textsc{sonic} calculations of the same quantity. 
Note that the data points shown include nonflow contributions, 
whose estimated magnitude is accounted for in the asymmetric 
systematic uncertainties.}
\label{fig:figure4}
\end{figure}

To further explore this idea, we divide the $v_2$ curves by their 
corresponding $\varepsilon_2$ from Table~\ref{table_geometry}, 
attempting to establish a scaling relation between the two quantities. 
Fig.~\ref{fig:figure4} shows that the ratios do not collapse to a common 
value.  As expected, this behavior is also reproduced by the \textsc{sonic} 
calculation, because both data and calculation are divided by 
the same $\varepsilon_2$ values. The lack of scaling in the 
\textsc{sonic} calculation can be understood from \dau events where the 
neutron and proton from the deuteron projectile are far separated and 
create two hot spots upon impacting the Au nucleus. These events have a 
large $\varepsilon_2$, but can result in small $v_2$ if the two hot 
spots evolve separately, never combining within the hydrodynamic time 
evolution. This effect is present in the \dau and \hau systems, and 
lowers the average $v_2/\varepsilon_2$ as detailed in 
Ref.~\cite{nagle_exploiting_2013}.

\begin{figure*}[htbp]
  \includegraphics[width=0.998\linewidth]{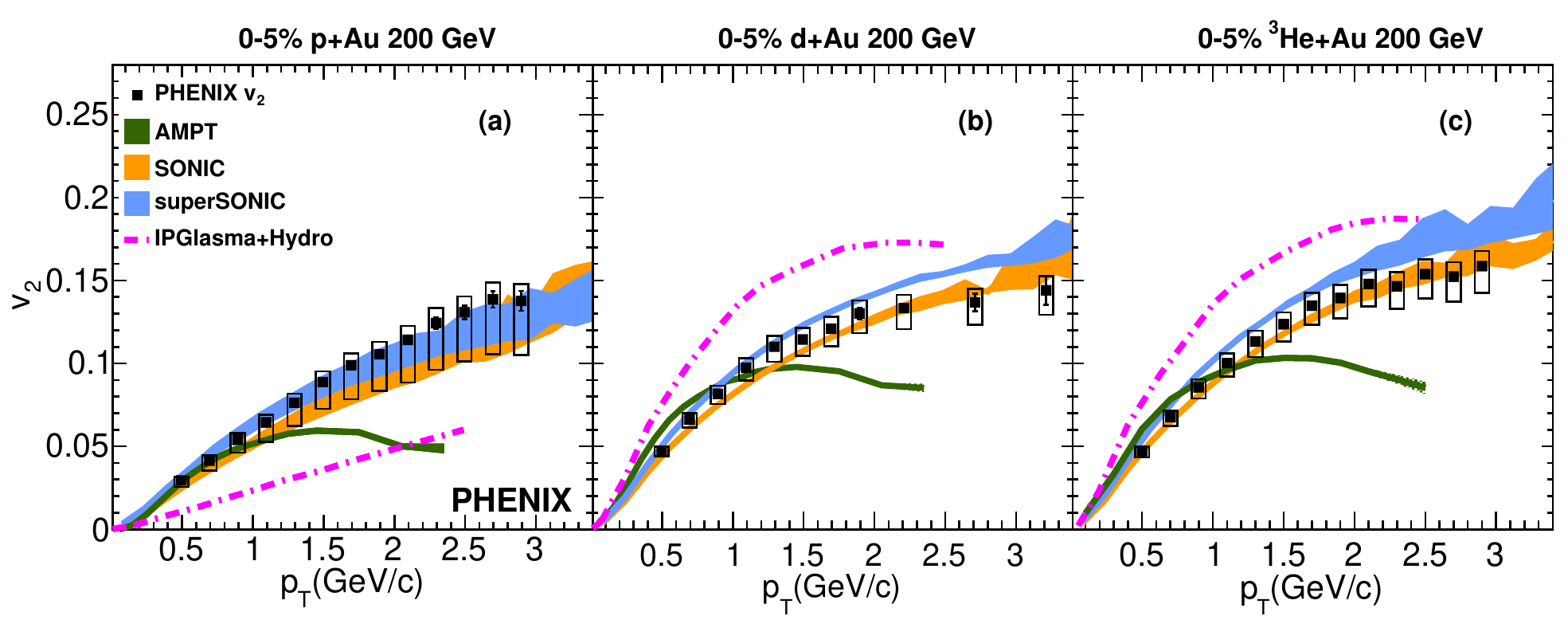}
\caption{Transverse momentum dependence of $v_2$ in central 0\%--5\%  
(a) \pau, (b) \dau, and (c) \hau collisions at \sqsn = 200 GeV. Theoretical 
calculations from (solid [gray] curve) \textsc{ampt}, 
(central [orange] band) \textsc{sonic}, (top [blue] band) \textsc{superSONIC}, 
and (dot-dashed [magenta] curves) IPGlasma+Hydro are shown in each panel. 
Note that the data points shown include nonflow contributions, 
whose estimated magnitude is accounted 
for in the asymmetric systematic uncertainties.}
\label{fig:figure5}
\end{figure*}

Figure ~\ref{fig:figure5} shows $v_2(\pt)$ for 0\%--5\% central \pau, 
\dau, and \hau events, along with theoretical predictions available in 
the literature, most notably from hydrodynamics with Glauber initial 
conditions (\textsc{sonic}~\cite{Habich:2014jna} and 
\textsc{supersonic}~\cite{Romatschke:2015gxa}), hydrodynamics with 
IP-Glasma initial conditions~\cite{Schenke:2014gaa}, and 
A-Multi-Phase-Transport Model 
(\textsc{ampt})~\cite{lin_multiphase_2005}.

The \textsc{supersonic} model uses the same prescription for initial 
conditions, hydrodynamic expansion, and hadronic cascade as 
\textsc{sonic}, yet additionally incorporates pre-equilibrium dynamics 
with a calculation in the framework of the AdS/CFT 
correspondence~\cite{vanderSchee:2013pia,Chesler:2015wra,Romatschke:2013re}. 
These two models agree well with the data within uncertainties, 
supporting the idea of initial geometry as the driver of the $v_n$ 
signal.  Furthermore, this illustrates how these results impose useful 
constraints to reduce the number of \emph{free parameters} of the model, 
because many such parameters must be identical across systems, e.g., 
$\eta/s$, the transition temperature to a hadron cascade, and the Monte 
Carlo Glauber smearing of nucleon coordinates of $\sigma=0.4$ fm.

Calculations using IP-Glasma initial conditions followed by viscous 
hydrodynamics have been successfully used to describe collectivity in 
A+A collisions~\cite{Gale:2012rq}. It is notable that in these 
calculations the glasma framework is used only to determine the initial 
spatial configuration as input to hydrodynamics; there is no glasma 
diagram or momentum-domain physics incorporated, such that all of the 
collectivity arises from final-state interactions. When this framework 
is applied to small collision systems with $\eta/s = 0.12$ and $b < 2$ 
fm, as shown in Fig.~\ref{fig:figure5}, the calculation substantially 
overestimates the data for \dau and \hau, while underestimating it for 
\pau. This follows from the fact that IP-Glasma generates very 
\emph{circular} initial conditions for \pau, corresponding to very low 
$\varepsilon_2$ values; however, the presence of several hot spots in 
\dau and \hau result in IP-Glasma values for $\varepsilon_2$ more 
comparable to those from Glauber. This is shown in 
Table~\ref{tab:glasma_geometry}.

\begin{table}[h!]
\caption{Initial eccentricity $\varepsilon_2$ of small systems at \sqsn 
= 200 GeV for 0\%--5\% centrality from Monte Carlo Glauber initial 
conditions smeared with a two-dimensional Gaussian of width $\sigma=0.4$ 
fm, and IP-Glasma initial conditions.}
\begin{ruledtabular}
\begin{tabular}{c c c c}
\label{tab:glasma_geometry}
 & \pau & \dau & \hau \\ \hline
 Glauber $\langle \varepsilon_2 \rangle$ & $0.23\pm 0.01$ & $0.54\pm 0.04$ & $0.50\pm 0.02$ \\
 IP-Glasma $\langle \varepsilon_2 \rangle$ & $0.10\pm 0.02$ & $0.59\pm 0.01$ & $0.55\pm 0.01$ \\
\end{tabular}
\end{ruledtabular}
\end{table}

In the case of \dau and \hau, a better agreement with data can be 
achieved by increasing the value of $\eta$/s or by including a hadronic 
cascade stage. However, doing so would lower the prediction for \pau 
even further. This demonstrates that IP-Glasma does not generate the 
appropriate initial conditions to account for measured $v_n$ via 
hydrodynamic flow.

It is important to notice that additional degrees of freedom for the 
geometry of \pau collisions arise from fluctuations of the shape of the 
proton, as described in Ref.~\cite{Schlichting:2014ipa}. The 
contribution of this effect to the measured elliptic flow may be 
constrained by $p$$+$$p$ data, and also possibly by varying the target in 
other $p+$A systems.

An additional framework accounting for subnucleonic degrees of freedom 
extends the Monte Carlo Glauber approach to also incorporate collisions 
between constituent quarks~\cite{Eremin:2003qn}. Recently, this 
framework has been successfully applied to the description of 
midrapidity charged particle multiplicity and transverse energy 
production~\cite{Adler:2013aqf,Adare:2015bua}. Different implementations 
of constituent quark Monte Carlo Glauber calculations are detailed in 
Refs.~\cite{Welsh:2016siu,Loizides:2016djv,Bozek:2016kpf,Mitchell:2016jio}. 
In Fig. 13(f) of Ref.~\cite{Welsh:2016siu}, the initial eccentricities 
$\varepsilon_2$ in \pau, \dau, and \hau obtained by incorporating 
constituent quarks in addition to multiplicity fluctuations are found to 
be $\varepsilon_2 =$ 0.42, 0.54, and 0.54, respectively. This 
calculation assumes a Gaussian density distribution of low-$x$ gluons 
around each constituent quark, of width $\sigma_g=0.3$ fm. It is 
interesting to note that the \dau and \hau systems show little 
sensitivity to the incorporation of both constituent quarks and 
multiplicity fluctuations into the calculation of the initial 
$\varepsilon_2$. Conversely, under the same circumstances, \pau has a 
substantially larger $\varepsilon_2$ than in the models shown in 
Table~\ref{tab:glasma_geometry}. Ref.~\cite{Welsh:2016siu} also 
presents calculations incorporating nucleonic degrees of freedom and 
multiplicity fluctuations, in which case a lower $\varepsilon_2=0.34$ is 
obtained for \pau. This shows that, when compared to the Glauber 
$\varepsilon_2$ for \pau in Table~\ref{tab:glasma_geometry}, 
quark-level degrees of freedom and multiplicity fluctuations may both 
play a significant role. Hydrodynamic calculations with these initial 
conditions will be of interest for future studies.

Finally, \textsc{ampt} combines partonic and hadronic scattering in a 
single model. Central \textsc{ampt} events with impact parameter $b<2$ 
have a midrapidity $dN_{ch}/d\eta$ = 8.1, 14.8, and 20.7 for \pau, \dau, 
and \hau, respectively. These were generated with the same Monte Carlo 
Glauber initial conditions used to characterize event geometry, and thus 
have very similar eccentricities to those given in Table I. Using the 
initial Glauber geometry information to compute $v_2$ relative to the 
participant plane~\cite{Koop:2015wea} yields results that agree 
reasonably well with the data below $\pt \approx 1$ GeV/$c$, yet 
underpredict them at higher \pt. It is noteworthy that despite the very 
different physics of \textsc{ampt} compared to the other models, it has 
successfully been applied to a variety of systems at RHIC and the LHC. 
See, for example, 
Refs.~\cite{Adare:2015cpn,Koop:2015wea,Ma:2016fve,Ma:2014pva}

\section{Summary}

We have presented results on azimuthal anisotropy and elliptic flow in 
central \pau at \sqsn = 200 GeV, compared with $v_2$ in \dau and \hau 
collisions. These results impose strong constraints on any model 
attempting to describe small system collectivity, whether by the 
formation of strongly interacting hot nuclear matter, or other 
mechanisms. We observe an imperfect scaling of $v_2$ with 
$\varepsilon_2$, well reproduced by hydrodynamics, providing strong 
evidence for initial geometry as the source of final-state momentum 
anisotropy in these systems. This disfavors other explanations based on 
initial-state momentum space domain effects. Further insight into the 
nature of small system collectivity can be gained by analyzing the 
centrality and collision energy dependence of $v_2$, and will be the 
subject of future studies.

\section*{Acknowledgments}

We thank the staff of the Collider-Accelerator and Physics
Departments at Brookhaven National Laboratory and the staff of
the other PHENIX participating institutions for their vital
contributions.  We acknowledge support from the
Office of Nuclear Physics in the
Office of Science of the Department of Energy,
the National Science Foundation,
Abilene Christian University Research Council,
Research Foundation of SUNY, and
Dean of the College of Arts and Sciences, Vanderbilt University
(U.S.A),
Ministry of Education, Culture, Sports, Science, and Technology
and the Japan Society for the Promotion of Science (Japan),
Conselho Nacional de Desenvolvimento Cient\'{\i}fico e
Tecnol{\'o}gico and Funda\c c{\~a}o de Amparo {\`a} Pesquisa do
Estado de S{\~a}o Paulo (Brazil),
Natural Science Foundation of China (People's Republic of~China),
Croatian Science Foundation and
Ministry of Science, Education, and Sports (Croatia),
Ministry of Education, Youth and Sports (Czech Republic),
Centre National de la Recherche Scientifique, Commissariat
{\`a} l'{\'E}nergie Atomique, and Institut National de Physique
Nucl{\'e}aire et de Physique des Particules (France),
Bundesministerium f\"ur Bildung und Forschung, Deutscher
Akademischer Austausch Dienst, and Alexander von Humboldt Stiftung (Germany),
National Science Fund, OTKA, K\'aroly R\'obert University College,
and the Ch. Simonyi Fund (Hungary),
Department of Atomic Energy and Department of Science and Technology (India),
Israel Science Foundation (Israel),
Basic Science Research Program through NRF of the Ministry of Education (Korea),
Physics Department, Lahore University of Management Sciences (Pakistan),
Ministry of Education and Science, Russian Academy of Sciences,
Federal Agency of Atomic Energy (Russia),
VR and Wallenberg Foundation (Sweden),
the U.S. Civilian Research and Development Foundation for the
Independent States of the Former Soviet Union,
the Hungarian American Enterprise Scholarship Fund,
and the US-Israel Binational Science Foundation.


%
 
\end{document}